\documentclass[slac_one]{revtex4}
\usepackage{epsfig}
\usepackage{subfigure,rotating,rotate}
\usepackage{graphicx}
\usepackage{fancyhdr}
\def\babar{\mbox{\slshape B\kern-0.1em{\small A}\kern-0.1em                             
    B\kern-0.1em{\small A\kern-0.2em R}}}
\def\pep2{PEP-II}

\newcommand{\Dz}{\ensuremath{D^0}}
\newcommand{\Dzb}{\ensuremath{\overline{D^0}}}
\newcommand{\Dzbar}{\ensuremath{\overline{D^0}}}

\newcommand{\DzDzb}{\Dz-\Dzbar}

\newcommand{\Kp}{K^+}
\newcommand{\Km}{K^-}
\newcommand{\KS}{K_S}

\newcommand{\pim}{\pi^-}
\newcommand{\pip}{\pi^+}
\newcommand{\piz}{\pi^0}

\newcommand{\bea}{\begin{eqnarray}}
\newcommand{\eea}{\end{eqnarray}}
\newcommand{\beq}{\begin{equation}}
\newcommand{\eeq}{\end{equation}}

\newcommand{\ycp}{\ensuremath{y_{\sst C\!P}}}
\newcommand{\CP}{\ensuremath{C\!P}}
\newcommand{\CPV}{\ensuremath{C\!PV}}
\newcommand{\ifb}{\ensuremath{fb^{-1}}}

\newcommand{\Dp}{\ensuremath{D^+}}

\newcommand{\epem}{\ensuremath{e^+e^-}}

\newcommand{\sst}{\scriptscriptstyle}

\begin{document}

\title{Mixing and $CP$ Violation in Decays of Charm Mesons} 

\author{B. Meadows
             \footnote{From the \babar\ collaboration.}}
\affiliation{U. Cincinnati, OH 45221, USA
}

\begin{abstract}
The phenomenon of mixing in neutral meson systems has now been observed in 
all flavours, but only in the past year in the $\Dz$ system.  The standard 
model anticipated that, for the charm sector, the mixing rate would be small, 
and also that \CP\ violation, either in mixing or in direct decay, would be 
below the present levels of observability.
It is hoped that further study of these phenomena might reveal signs of new
physics.  A review of recently available, experimental results is given.
\end{abstract}

\preprint{UCHEP-08-06}
\maketitle

\thispagestyle{fancy}

\section{INTRODUCTION\label{sec:intro}}
Mixing and \CP\ violation (\CPV) in the neutral $D$ system were first
discussed over thirty years ago 
\cite{Pais:1975qs}
but mixing was observed for the first time only very recently
\cite{Aubert:2007wf,Staric:2007dt}.
Since then, these observations have been confirmed
\cite{Abe:2007rd,Aaltonen:2007uc,Aubert:2007en,Aubert:2008zh}
in other experiments and in other $\Dz$ decay modes.

Unlike the $K$, $B$ and $B_s$ systems, for which mixing was observed years
earlier, the short distance $(\Delta C=2)$ amplitude contributing to mixing
in the $D$ system arises from box diagrams with down- rather than up-type
quarks in the loops.  The $d$ and $s$ components are GIM-suppressed, and
the $b$ component is suppressed by the small $V_{ub}$ CKM coupling.  In
the standard model (SM), therefore, long range, non-perturbative effects,
a coherent sum over intermediate states accessible to both $\Dz$ 
and $\Dzbar$, are the main contribution to mixing.  These are hard to
compute reliably, however
\cite{Wolfenstein:1985ft,Bigi:2000wn,Golowich:2007ka,Golowich:2006gq}.

\subsection{Notation and Formalism\label{sec:formalism}}
\Dz\ and \Dzbar\ mesons are produced in flavour eigenstates, but 
propagate in time $t$ and decay as mixtures of eigenstates $D_1$ 
and $D_2$, with masses and widths $M_{1,2}$ and $\Gamma_{1,2}$,
related to the flavour states by:
\begin{eqnarray}
  \label{eq:eigenstates}
  \begin{array}{ccccccc}
     |D_1\!\!> &\!\!=\!\!& p|\Dz\!\!>\!+ q|\Dzbar\!\!> &
     \quad ~~;~~
     |D_2\!\!> &\!\!=\!\!& p|\Dz\!\!>\!- q|\Dzbar\!\!> 
  \end{array}
\end{eqnarray}
It is usual to define mixing parameters $x$, $y$, $r_{\sst M}$ and 
$\phi_{\sst M}$ and a decay parameter $\lambda_f$:
\begin{eqnarray}
  \label{eq:notations}
  x  = {M_1-M_2\over\Gamma}  ~~;~~
  y  = {\Gamma_1-\Gamma_2\over 2\Gamma}  ~~;~~
  r_{\sst M}=\left|{q\over m}\right|  ~~;~~
  \phi_{\sst M}={\rm Arg}\left\{{q\over p}\right\}  ~~;~~
  \lambda_f =  {q\bar{\cal A}_f \over p{\cal A}_f} 
  \propto  e^{i(\delta_f+\phi_f)}
\end{eqnarray}
in which $M$ and $\Gamma$ are averages for the two mass eigenstates
and $\delta_f$, $\phi_f$ are strong and weak relative phases, 
respectively, for the amplitudes ${\cal A}_f(\bar{\cal A}_f)$ for 
the decays of $\Dz(\Dzbar)$ to the final state $f$.  In the absence 
of \CPV\ in mixing
$p=q=1/\sqrt 2$ and $\phi_M$ (the mixing phase) is zero, $D_1$ is
\CP-even and $D_2$ is \CP-odd.  If there is no direct \CPV\ in
decay, then either $\phi_f$ or $\delta_f=0$.

Recent estimates for $|x|\sim 1$\%, $|y|\sim 1$\% 
are in good agreement with the present observations and are orders of 
magnitude smaller than values measured for the other neutral mesons.  
Since charm decays are dominated by the $2\times 2$ sector of the CKM
matrix, where weak phases are absent, \CPV, in either mixing,
or in time-integrated decay rates, is unexpected.  Either a large
value for $|x|$ (e.g. $>|y|$) or the observation, at current 
sensitivities, of \CPV\ in charm decays would be regarded as an
indication of new physics
\cite{Bigi:2000wn,Golowich:2007ka,Golowich:2006gq,Grossman:2006jg}.

In the following section, we discuss the experimental evidence for 
mixing as observed in time-dependent effects in hadronic decays.  
Then, in section~\ref{sec:lifetimes} we present results from 
its effect on lifetimes.  In section \ref{sec:cpv}, we
outline the present state of experimental 
searches for\CPV\ in time-integrated decay rates of charmed
mesons in section~\ref{sec:cpv}.  In section~\ref{sec:summary},
we present a summary of the current knowledge on mixing parameters
and outline possibilities for future progress.

\section{EVIDENCE FOR MIXING IN TIME-DEPENDENCE OF DECAYS \label{sec:timedep}}
It follows from Eqs.~(\ref{eq:eigenstates}) and (\ref{eq:notations})
that mixing gives rise to a time-dependence in amplitudes
${\cal A}(\bar{\cal A})_{f(\bar f)}$ for decays $\Dz(\Dzbar)\to f(\bar f)$.
When the final state $f$ is accessible to decay by both $\Dz$
and $\Dzbar$, interference between direct decay and decay preceded
by mixing will occur.  Neglecting \CPV\ and terms in $x$ or $y$
beyond second order, the decay rate for $\Dz\to\bar f$, for example,
deviates from exponential
\begin{eqnarray}
  \label{eq:wsrs}
   |{\cal A}_{\bar f}(t)|^2 = e^{-\Gamma t}\times
       \left[|{\cal A}_{\bar f}|^2 + 
             \left(y\cos\delta_f-x\sin\delta_f\right)
             |{\cal A}_{\bar f}||\bar{\cal A}_{\bar f}|
             (\Gamma t) +
             {(x^2+y^2)\over4}
             |\bar{\cal A}_{\bar f}|^2
             (\Gamma t)^2
             \right].
\end{eqnarray}
The first term in square parentheses is the direct decay rate, the
third is proportional to the mixing rate $(x^2+y^2)/2$ and the middle
term comes from the interference.  The latter makes it possible to
measure the small quantities $x$ and $y$ to first order, but only if
the strong phase difference $\delta_f$ is known.

\subsection{Evidence for Mixing In $\mathbf{\Dz\to\Kp\pim}$ Decays}
Observation of the deviations from exponential in the ``wrong-sign"
(WS) decays $\Dz\to\Kp\pim$
\footnote{Hereafter, unless otherwise stated, it is implied that
charge conjugate modes are included.},
a signal for $\DzDzb$ mixing, has been attempted many times
\cite{Aitala:1996fg,Godang:1999yd,Link:2004vk,Abe:2004sn,
 Aubert:2006rz,Zhang:2006dp}
without success.  These very rare decays are Doubly-Cabibbo-suppressed
(DCS) and must be compared, experimentally, with the far more copious,
``right-sign" (RS) Cabibbo-Favoured (CF) decays to $\Km\pip$, in
which the deviations are negligibly small.

The \babar\ collaboration, using a huge 384\ifb\ \epem\ sample of
$\Dz$ mesons, whose flavour was determined from the sign of
the pion from the $D^{\ast+}\to\Dz\pip$ decays from which they came,
were finally able to observe mixing.  In Fig.~\ref{fig:evidence}(a),
the ratio of WS to RS decays of the combined sample of $\Dz$ and 
$\Dzbar$ decays is shown as a function of proper time $t$.  The 
average value of this ratio, $\sim 0.36$\%, is clearly not constant.  
The mixing fit, including a linear rise corresponding to the second 
term in Eq.~(\ref{eq:wsrs}), is preferred.
 \begin{figure}[h]
 \begin{center}
 \begin{tabular} {ccc}
  \subfigure[]{\epsfig{file=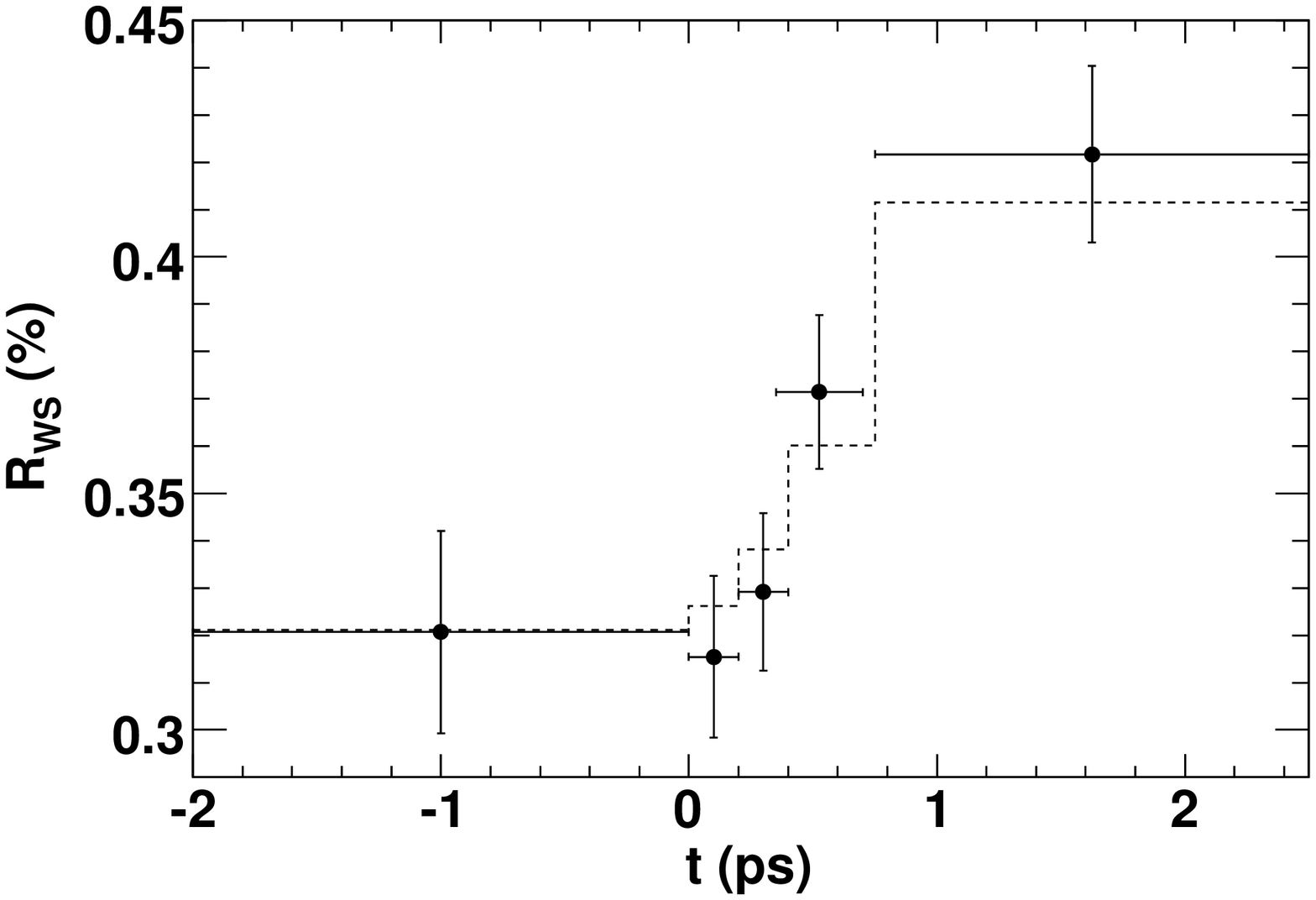,height=0.20\linewidth,width=0.22\linewidth}} &
  \hspace{0.1\linewidth}
  \subfigure[]{\epsfig{file=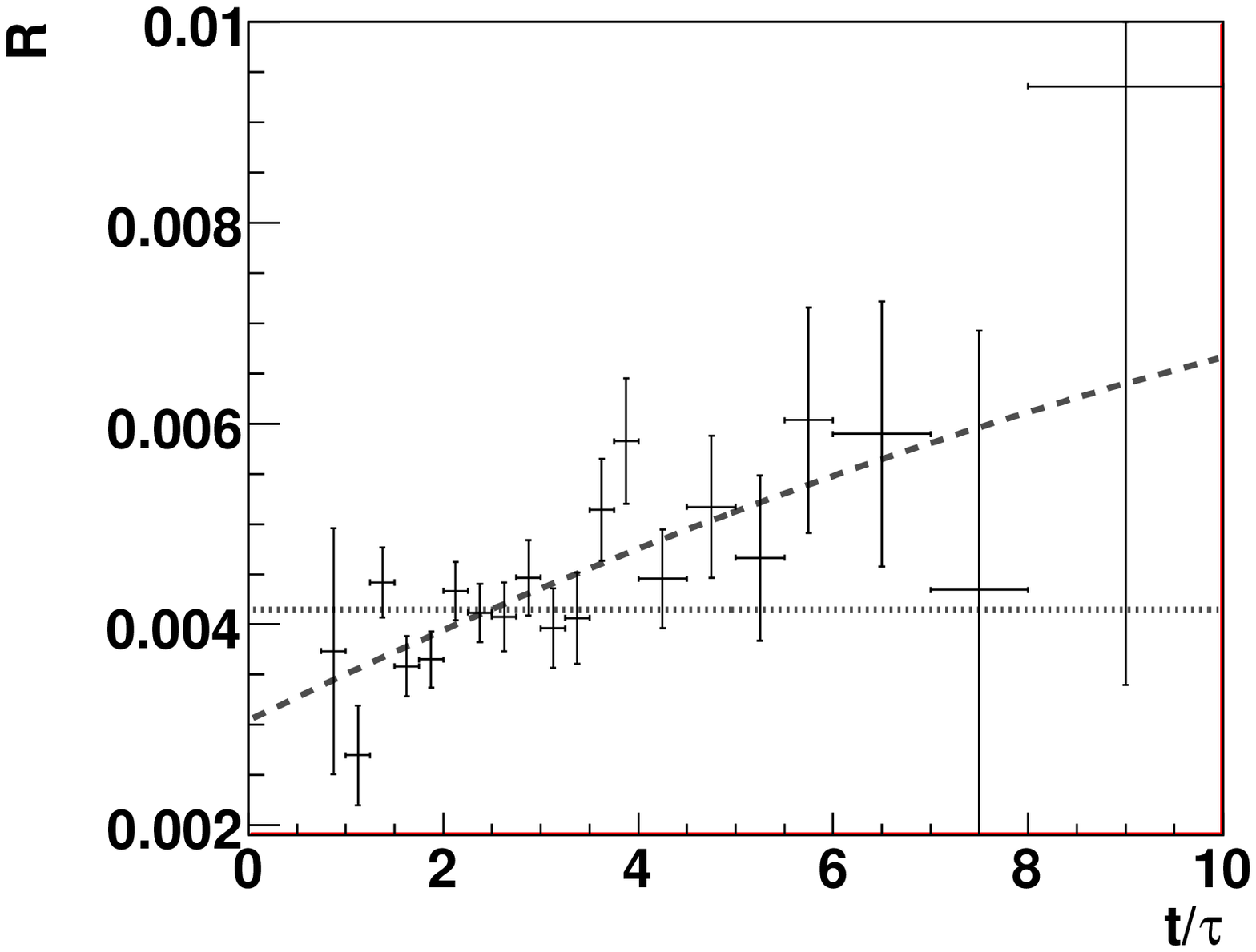,height=0.20\linewidth,width=0.22\linewidth}}
  \hspace{0.1\linewidth}
  \subfigure[]{\epsfig{file=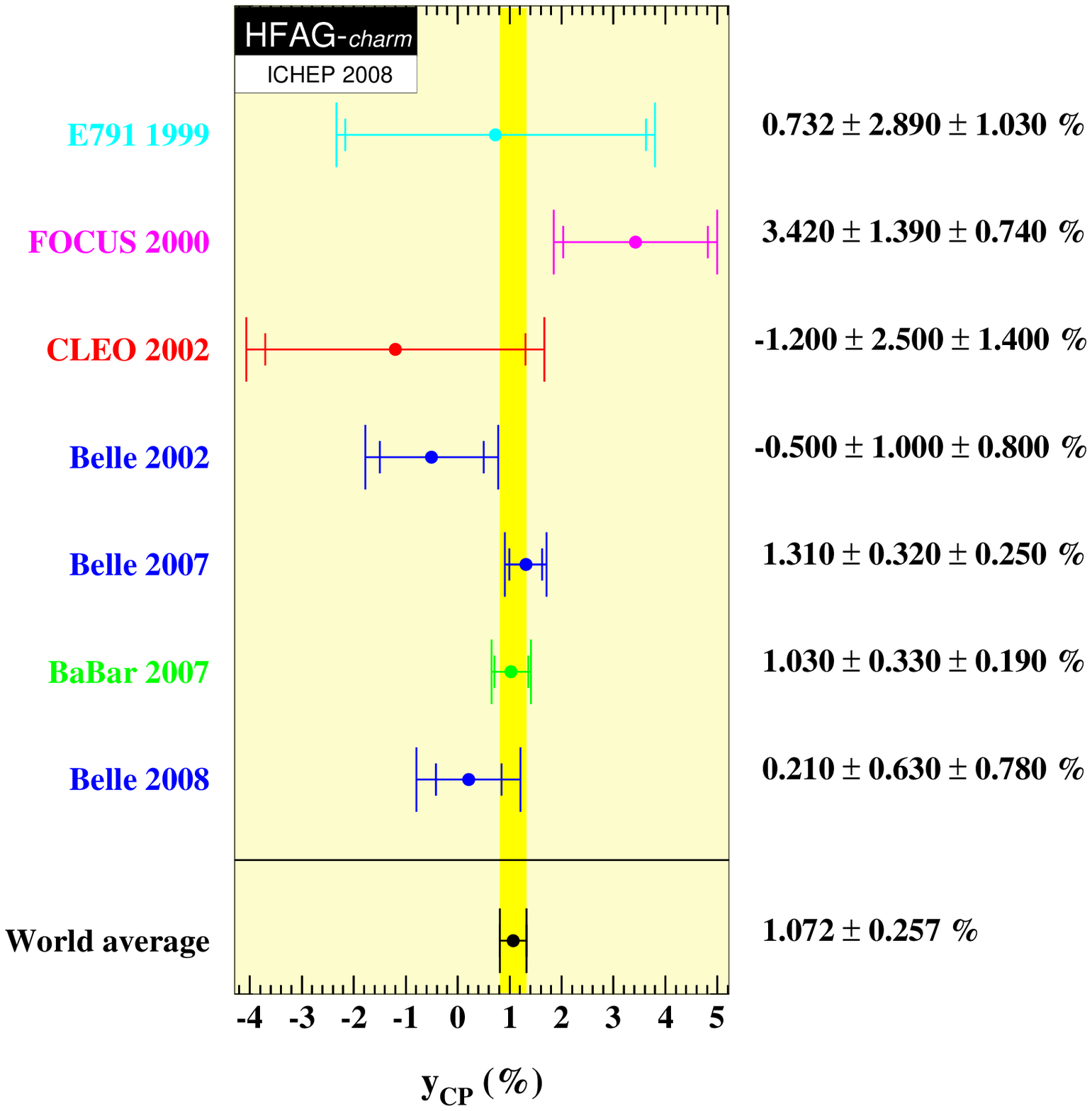,width=0.22\linewidth,height=0.20\linewidth}} &
 \end{tabular}
 \end{center}
 \caption{\label{fig:evidence}
 Evidence for $\DzDzb$ mixing.  The ratio of WS/RS yields as a function
 of proper time $t$ is shown for the (a) \babar\ and (b) CDF experiments.  
 The dashed plot in (a) shows the projection of the mixing fit onto the
 time axis, and has a $\chi^2$ of 1.5.  The flat line (no mixing fit) 
 has $\chi^2$ of 24.
 The dashed line in (b) is a parabolic fit giving values for the DCS 
 rate, $x'^2$ and $y'$.
 In (c), the Heavy Flavour Averaging Group (HFAG)
 \cite{hfag:ycp}
 summary of measurements of \ycp\ is presented.  The mean,
 $1.072\pm 0.257$\%, differs significantly from the no-mixing
 expectation of zero.
 }
 \end{figure}

The strong phase $\delta_{K\pi}$ was poorly known, so this measurement
was only able to determine values for $y'=(9.7\pm 4.4\pm 3.1)\times 10^{-3}$,
$x'^2=(-0.22\pm 0.30\pm 0.21)\times 10^{-3}$
and $R_{\sst D}=(0.303\pm 0.019)$\%, where
\[  x'=x\cos\delta_{K\pi}+y\sin\delta_{K\pi}~~;~~
    y'=y\cos\delta_{K\pi}-x\sin\delta_{K\pi}
\]
and $R_{\sst D}$ is the ratio of DCS ro CF decay rates.
The \babar\ result was found to be inconsistent with a no mixing hypothesis, 
with a significance estimated at $3.9\sigma$.
These results assumed no \CPV.  Separate fits to \Dz\ and to \Dzbar\
showed no significant difference.  It was concluded that
no \CPV\ was observable at the current precision.

Using a similar technique, but with data from 1.96~TeV $p\bar p$
collisions, the CDF experiment has confirmed this result with a
$3.8\sigma$ significance
\cite{Aaltonen:2007uc}.
The plot of WS/RS ratio vs. $t$ is shown in Fig.~\ref{fig:evidence}(b).
Assuming no \CPV, they obtained values
$x'^2=(-0.12\pm 0.35)\times 10^{-3}$, $y'=(8.5\pm 7.6)\times 10^{-3}$
and $R_{\sst D}=(0.304\pm 0.055)$\%,
in excellent agreement with those from \babar.

\subsection{Lifetimes for Decays to \CP-even Final States\label{sec:lifetimes}}
Deviations from exponential decay, introduced by mixing, alters the
effective lifetimes $\tau$ for $\Dz$ decays in ways that depend upon the \CP\
symmetry of the final state
\cite{Liu:1994ea,Bergmann:2000id}.
Decays to $\Km\pip$ (mixed \CP) and to $\Km\Kp$ or $\pim\pip$ (\CP-even)
have been used in searches for mixing by measuring the quantities
\bea
  \ycp = {\tau(\Dz\!\!\to\!K^-\pi^+)\over
                 \tau(\Dz\!\!\to\! h^-h^+)} - 1
  ~~;~~
  A_{\tau} = 2\tau_{\Km\pip}\left({\tau^+_{hh}-\tau^-_{hh}\over
                              \tau^+_{hh}+\tau^-_{hh}}\right)
\eea
Both are zero in the absence of mixing or \CPV.  Clear evidence for
mixing ($\ycp\ne 0$) has now been seen in this way by both the Belle
\cite{Staric:2007dt} and \babar\ 
\cite{Abe:2007rd}
collaborations with significances of $3.2\sigma$ and $3.0\sigma$,
respectively
These results, with those of earlier, unsuccessful
searches for mixing, are summarized in Fig.~\ref{fig:evidence}.

No evidence for \CPV\ ($A_{\tau}\ne 0$) is yet observed, however.  
Belle measured $A_{\tau}=(0.010\pm 0.300\pm 0.150)$\% and
\babar\ obtained
\footnote{\babar\ measured the asymmetry in 
 $\Delta Y=(\ycp^+-\ycp^-)/(\ycp^++\ycp^-)\approx 2\tau^0 
 A_{\tau}/(\tau^++\tau^-)$.}
$A_{\tau}=(0.260\pm 0.360\pm 0.080)$\% giving an average
$A_{\tau}=(0.123\pm 0.248)$\%.

\subsection{Mixing in $\mathbf{\Dz\to\Kp\pim\piz}$ Decays}
The \babar\ experiment has just reported further evidence for 
mixing in WS $\Dz$ decays to $\Kp\pim\piz$.
\cite{Aubert:2008zh}.
For this three-body channel, each decay is represented as a point
in a Dalitz plot, defined by the squared invariant masses 
$(s_0,s_+)$ of the $\Kp\pim$ and $\Kp\piz$ systems, respectively.
The final state $f$ and the value for the strong phase 
$\delta_f=\delta_{K\pi\pi} +
        {\rm Arg}\left(\bar{\cal A}_{\bar f}/
                        {\cal A}_{\bar f}\right)$
in Eq.~(\ref{eq:wsrs}) is then unique at each such point.  The
CF and DCS decay amplitudes $\bar{\cal A}_{\bar f}$ and
${\cal A}_{\bar f}$, respectively, can be determined from
fits to their Dalitz plot densities, but the constant phase 
$\delta_{K\pi\pi}$ cannot.  

The \babar\ collaboration identified a sample of $\sim 3,000$ WS
and $\sim 660,000$ RS $\Dz\to K\pi\piz$ events in a tight signal 
region with flavours identified by their origin from $D^{\ast+}$ 
decays.  The RS sample was very clean, though the WS sample had a
50\% background, mostly from ``mis-tagged" RS decays.
  
An isobar model for the amplitude $\bar{\cal A}_{\bar f}$
was determined from a time-integrated fit to the distribution
of events on the RS (mostly CF) Dalitz plot.  A second,
time-dependent fit to the WS Dalitz plot then determined 
parameters for the DCS amplitude ${\cal A}_{\bar f}$ and
$x''$, $y''$ ($x$ and $y$ rotated by the unknown phase 
$\delta_{K\pi\pi}$) summarized in Table~\ref{tab:kpipi}.

\begin{table}[h]
 \begin{center}
 \caption{Mixing parameters
  $x^{\prime\prime}=x\cos\delta_{K\pi\pi}+y\sin\delta_{K\pi\pi}$
  and
  $y^{\prime\prime}=y\cos\delta_{K\pi\pi}-x\sin\delta_{K\pi\pi}$
  from the \babar\ analysis of $\Dz\to\Kp\pim\piz$ decays.  These
  are related to $x$ and $y$ by an undetermined strong phase
  rotation $\delta_{K\pi\pi}$.  
  }
 \begin{tabular}{|c|c|c|}
 \hline
  \textbf{Sample} &
  $\mathbf{x^{\prime\prime}}$ &
  $\mathbf{y^{\prime\prime}}$
 \\
 \hline
  $\mathbf{\Dz}$ \textbf{and} $\mathbf{\Dzbar}$ &
  $[~2.61^{+0.57}_{-0.68}\hbox{(stat.)}\pm 0.39\hbox{(syst.)}]$\% &
  $[-0.06^{+0.55}_{-0.64}\hbox{(stat.)}\pm 0.34\hbox{(syst.)}]$\%
 \\
  $\mathbf{\Dz}$ \textbf{only} &
  $[~2.53^{+0.54}_{-0.63}\hbox{(stat.)}\pm 0.39\hbox{(syst.)}]$\% &
  $[-0.05^{+0.63}_{-0.67}\hbox{(stat.)}\pm 0.50\hbox{(syst.)}]$\%
 \\
  $\mathbf{\Dzbar}$ \textbf{only} &
  $[~3.55^{+0.73}_{-0.83}\hbox{(stat.)}\pm 0.65\hbox{(syst.)}]$\% &
  $[-0.54^{+0.40}_{-1.16}\hbox{(stat.)}\pm 0.41\hbox{(syst.)}]$\%
 \\
 \hline
 \end{tabular}
 \label{tab:kpipi}
 \end{center}
\end{table}
The significance of the mixing signal in the fit to the
combined $\Dz$ and $\Dzbar$ samples, ignoring \CPV, was
$3.2\sigma$.  Separate fits to each sample show no evidence 
for \CPV.

\section{OTHER MIXING MEASUREMENTS}
Two other channels have recently been analyzed in which the 
mixing signal was of less significance, but where mixing 
parameters were, nevertheless, measurable.

\subsection{$\mathbf{\Dz\to\KS\pim\pip}$ Decays}
Analysis of this channel, pioneered by the CLEO collaboration
\cite{Asner:2005sz}
is of great interest due to the presence of \CP-eigenstates
in the final state.  These tie the phase of the amplitudes
for $\Dz$ and $\Dzbar$ decay amplitudes together, so that
$\delta_{K\pi\pi}$ in Eq.~(\ref{eq:deltakpipi}) is zero.
Thus, measurements of $x$, $y$ and the two \CPV\
parameters $|q/p|$ and $\phi_{\KS\pim\pip}$ can be made
free of unknown phases.

CLEO's analysis used a $9\ifb$ \epem\ sample and
found only an upper limit for mixing parameters
$(-4.7 < x < 8.6)$\% and $(-6.1 < y < 3.5)$\% at 95\% CL.  
The Belle collaboration has repeated the analysis with
a $540\ifb$ sample (534K events) 
\cite{Abe:2007rd},
observing mixing at the $2.4\sigma$ significance.  
The parameter values obtained are summarized in 
Table~\ref{tab:kspipi}.  It is noteworthy that the
value for $y$ here is not consistent with the world average 
value for $\ycp=1.072\pm 0.275$\% discussed in 
section~\ref{sec:lifetimes}. 

\begin{table}[h]
 \begin{center}
 \caption{Mixing parameters obtained from the Belle analysis of
  $\Dz\to\KS\pim\pip$ decays.  The third error is that due to
  uncertainties in the model used to define the Dalitz plot
  amplitudes.}
 \begin{tabular}{|c|c|c|c|}
 \hline
  $\mathbf{x}$     & 
  $\mathbf{y}$     & 
  $\mathbf{|q/p|}$ & 
  $\mathbf{{\rm Arg}(q/p)}$
 \\
 \hline
 $(0.80\pm 0.29^{+0.09 +0.10}_{-0.07 -0.14})$\% &
 $(0.33\pm 0.24^{+0.08 +0.06}_{-0.12 -0.08})$\% &
 $ 0.86^{+0.30 +0.06}_{-0.29 -0.03}\pm 0.08 $   &
 $(-14^{+16 +5 +2}_{-18 -3 -4})^{\circ}$
 \\
 \hline
 \end{tabular}
 \label{tab:kspipi}
 \end{center}
\end{table}
There was no evidence for \CPV\ since the magnitude of $q/p$
was consistent with unity and its phase was zero.

\subsection{\CP-odd Final States}
A new result from Belle 
\cite{phi:2008tp}.
has examined lifetime differences between the decays to the
\CP-odd final state $\Dz\to\phi\KS$ with decays to the \CP-even
background and $\phi$ sideband regions.  A novel method was 
chosen over a time-dependent analysis of the whole $\KS\Km\Kp$ 
Dalitz plot like that used for the $\KS\pim\pip$ channel which 
required a flavour-tagged sample of $\Dz$'s from $D^{\ast}$ 
decays.  The sample size available was thereby enhanced almost 
threefold. 

Approximately 130K $\Dz\to\Km\Kp\KS$ decays were extracted from
a 697\ifb\ sample of \epem\ collisions, with no attempt to
distinguish $\Dz$ from $\Dzbar$.  
Lifetimes $\tau_{\sst A}$ and $\tau_{\sst B}$ were measured,
respectively, for events in the $\Km\Kp$ invariant mass
regions A (containing the $\phi$) and B (narrow sidebands above
and below the $\phi$ where \CP-even $a_0(980)\KS$, $f_0(980)\KS$ 
and non-resonant decays were dominant).

Decay amplitudes 
${\cal A}_{1,2}={\cal A}(s_0,s_+)\pm{\cal A}(s_0,s_-)]/\sqrt 2$
were obtained from a fit
\cite{Aubert:2005sm,Aubert:2008bd}
to the distribution of events in the $\Km\Kp\KS$ Dalitz plot.
Here, $s_0$, $s_-$ and $s_+$ were, respectively, $\Km\Kp$, $\KS\Km$
and $\KS\Kp$ squared invariant mass combinations.  Neglecting 
\CPV, $|{\cal A}_{1,2}|^2$ should have approximately exponential 
time-dependences with lifetimes $\tau_{1,2}=\tau/(1\pm\ycp)$, 
and their interference ${\cal A}_1^*{\cal A}_2$ should be zero 
when integrated over the $s_+$ coordinate.

A value for \ycp\ was extracted from the difference between
$(\tau_{\sst A}$ and $\tau_{\sst B})$ and the ``\CP-ness", of
each region, obtained by integration of these amplitudes.

The result, $\ycp=(0.21\pm 0.63\pm 0.78\pm 0.01$, where the third
uncertainty accounts for uncertainties in the decay model used to
describe the decay amplitude ${\cal A}$, was included in 
Fig.~\ref{fig:evidence}(c).  This result ranks third in precision for
measurements of $\ycp$.  It is important that it agrees quite well
with all other results from the \CP-even modes.

\section{SEARCHES FOR \CPV\ IN CHARM MESON DECAYS \label{sec:cpv}}
SM expectations for \CP\ asymmetries in time-integrated decays in
the charm sector are small.  Singly-Cabibbo-Suppressed (SCS)
decays are of particular interest since these allow for asymmetries,
at the $\sim 0.1$\% level, from gluonic penguins
\cite{Buccella:1994nf,Bianco:2003vb,Petrov:2004gs,Grossman:2006jg}.
Observation of \CPV\ at current sensitivities would be an indication
for new physics.

The asymmetry is defined as
\[ A_{\sst CP}^f = {\Gamma(D\to f)-\Gamma(\bar D\to\bar f)\over
                    \Gamma(D\to f)+\Gamma(\bar D\to\bar f)}
\]
For $\Dz$ asymmetries, there could be contributions from both
direct \CPV\ and from mixing.  The latter is small since
$A_{\tau}\sim 0.1$\%.  

Experimentally, measurement of asymmetries at the $10^{-3}$ level
are limited by uncertainties in asymmetries in the detection
and reconstruction of particles of opposite charge.  Also, for
$\Dz$ decays, efficiencies for $D^{*+}$ tagging cannot be
assumed to be the same as that for $D^{*-}$.  Forward-backward
production asymmetries, resulting from $Z^0/\gamma$
interference and higher order loops in the production of
$c\bar c$ quarks, results in asymmetries in the distribution
of $D$ decay products in regions of varying efficiency in
the detector.  

Calibration of these factors used to rely upon Monte Carlo
simulated event (MC) studies, with questionable assumptions
about charge-dependent interaction effects, resulting in
systematic uncertainties in $A_{\sst CP}$'s in the $1-5$\%
range.  In the past year, new insights in using data rather
than MC have led to reduction of these uncertainties to
the $0.2-0.4$\% range
\cite{Aubert:2007if}.

Large samples of CF $\Dz\to\Km\pip$ decays, selected from
the \babar~$386\ifb$ data sample, were used to provide
information on the efficiency asymmetries in small momentum
and angular range.  The underlying assumption was that the
integrated production rates for $\Dz$ and $\Dzbar$ were the
same.  Forward-backward production asymmetries were dealt
with by measuring event yields in forward and backward
separately.

Asymmetries obtained in this way, are summarized in
Table~\ref{tab:cpv}.  Other, recent measurements are also included.

\begin{table}[h]
 \begin{center}
 \caption{Recent measurements of \CP\ decay asymmetries $\times 10^{-2}$.
  The first uncertainty is statistical, the second is systematic.
  Data were used to estimate asymmetries in efficiency in all but 
  ref.~\cite{Arinstein:2008zh}.}
 \begin{tabular}{|l|c|c|c|}
 \hline
  \textbf{Mode}   & 
  \textbf{\babar}
  ~~$\times 10^{-2}$
   &
  \textbf{Belle} 
  ~~$\times 10^{-2}$
   &
  \textbf{CLEO}
  ~~$\times 10^{-2}$
 \\
 \hline
   $\mathbf{\Dz\to\Km\Kp}$
   &
   $\:~0.00\pm 0.34 \pm 0.13$
   \cite{Aubert:2007if}
   &
   $\:-0.43\pm 0.30 \pm 0.11$
   \cite{cpv:2008rx}
   & ---
   \\
   $\mathbf{\Dz\to\pim\pip}$
   &
   $-0.24\pm 0.52 \pm 0.22$
   \cite{Aubert:2007if}
   &
   $~0.43\pm 0.52 \pm 0.12$
   \cite{cpv:2008rx}
   & ---
   \\
   $\mathbf{\Dz\to\Km\Kp\piz}$
   &
   $\:~1.00\pm 1.67 \pm 0.25$
   \cite{Aubert:2008yd}
   & --- 
   & ---
   \\
   $\mathbf{\Dz\to\pim\pip\piz}$
   &
   $-0.31\pm 0.41 \pm 0.17$
   \cite{Aubert:2008yd}
   &
   $-0.43\pm 0.41 \pm 1.23$
   \cite{Arinstein:2008zh}
   & ---
   \\
   $\mathbf{\Dp\to\Km\Kp\pip}$
   & ---
   & --- 
   &
   $-0.03\pm 0.84 \pm 0.29$
   \cite{cpv:2008zi}
   \\
 \hline
 \end{tabular}
 \label{tab:cpv}
 \end{center}
\end{table}
These asymmetries are consistent with zero, with most systematic
uncertainties much less than 1\%.  Of particular interest
is the observation that both statistical and systematic
uncertainties should scale with the square root of the number
of events.  Future measurements from flavour factories may well
detect asymmetries at or above the SM limit.

\section{SUMMARY \label{sec:summary}}
The Heavy Flavour Averaging Group (HFAG) has combined a wide range
of mixing observables (28 in all) including all those discussed 
here.  To do this, they made a $\chi^2$ fit
\cite{Schwartz:2008wa}
to obtain values for the underlying mixing parameters 
($x$, $y$, $|q/p|$,{\sl etc.}) that best described these.
The results are best summarized in the $\chi^2$ countour plots 
in Fig.~\ref{fig:mix_hfag} for the fit that allowed for \CPV.
 \begin{figure}[h]
 \begin{center}
 \begin{tabular} {cc}
  \subfigure[]{\epsfig{file=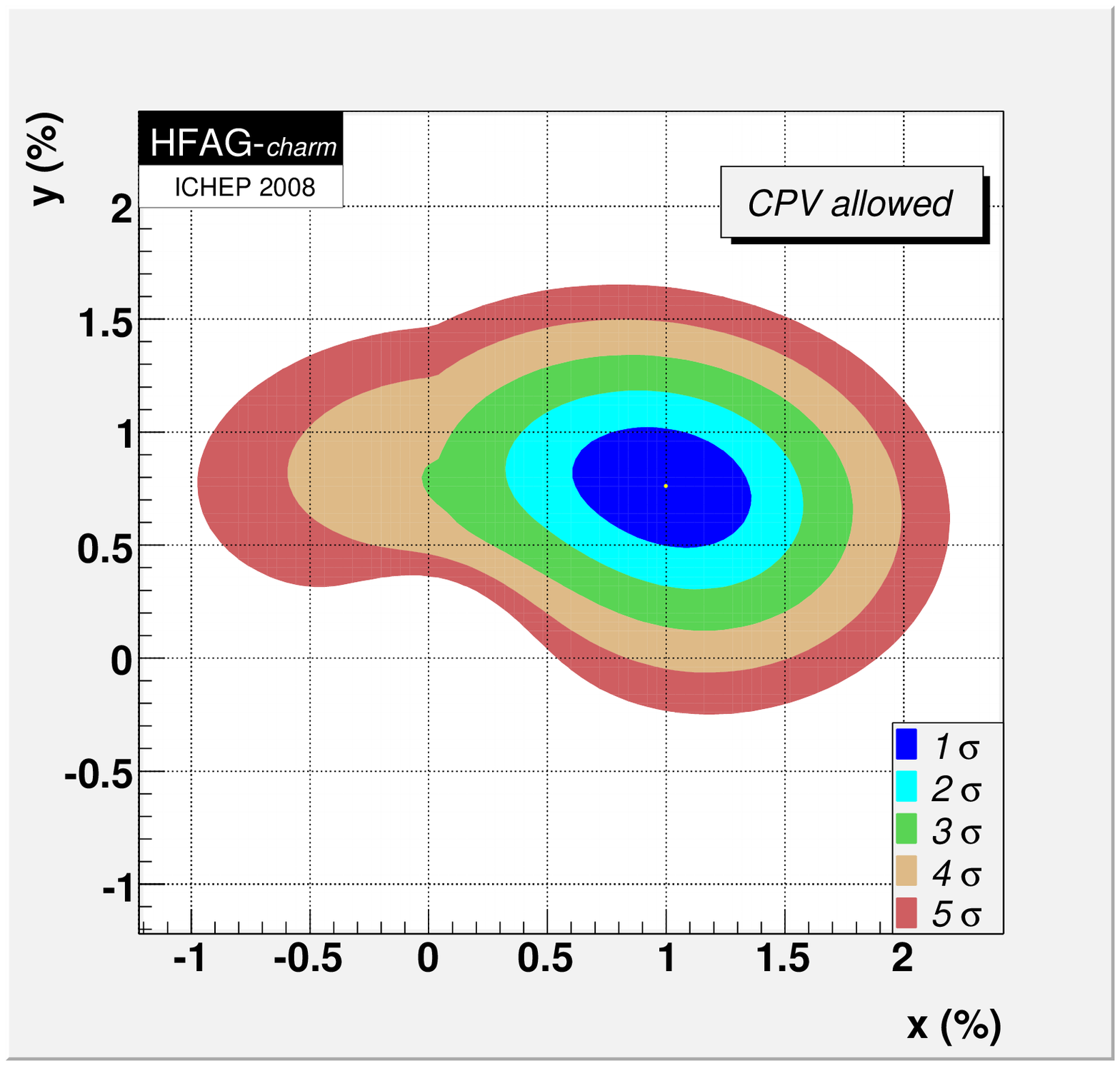,width=0.3\linewidth,height=0.20\linewidth}} &
  \hspace{0.1\linewidth}
  \subfigure[]{\epsfig{file=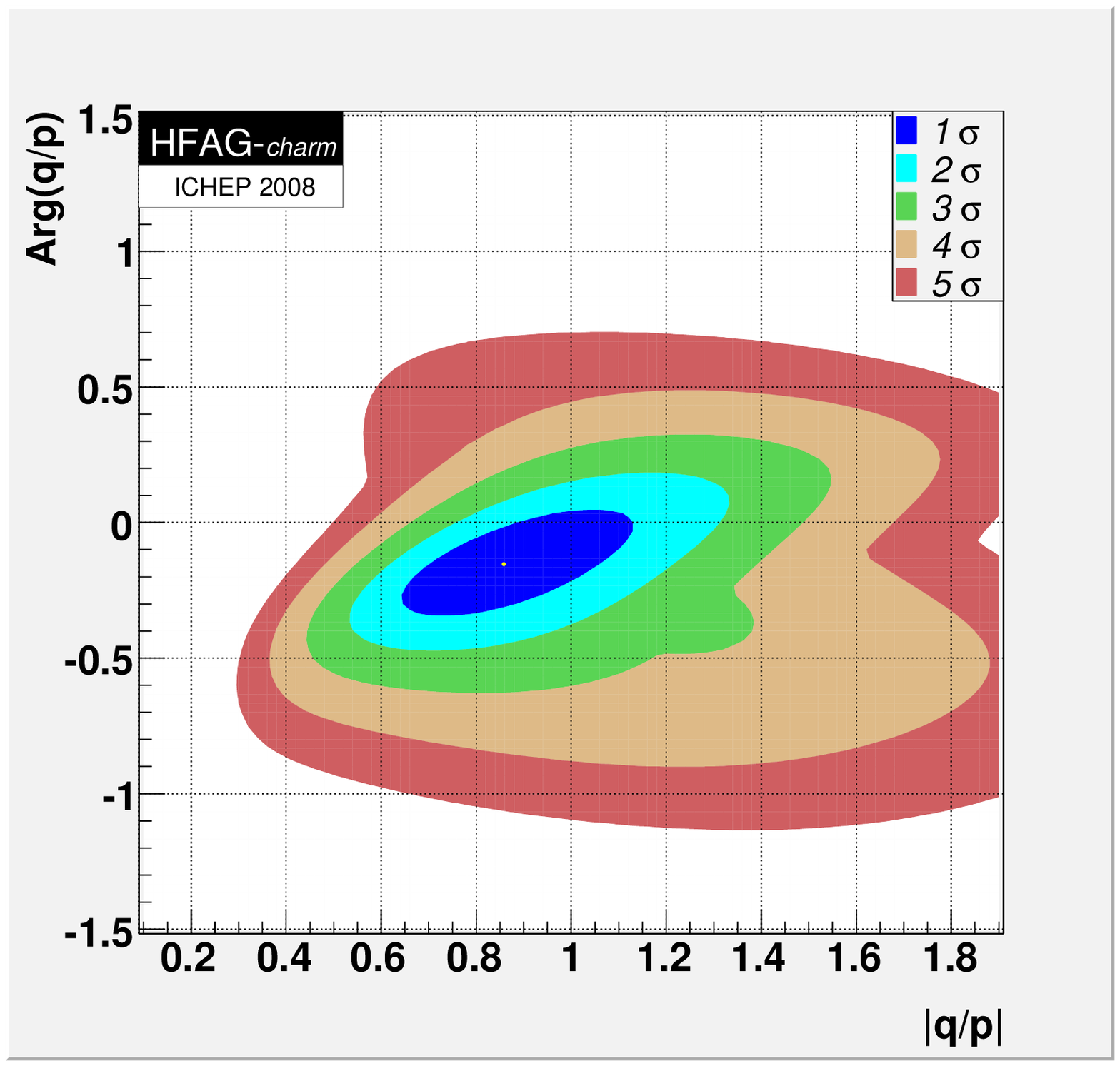,width=0.3\linewidth,height=0.20\linewidth}}
 \end{tabular}
 \end{center}
 \caption{\label{fig:mix_hfag}
  $\chi^2$ contours for fit to mixing observables. (a) $x$ vs. $y$
  and (b) $|q/p|$vs. Arg$(q/p)$.
  This figure is taken from the Heavy Flavour Averaging Group (HFAG)
 \cite{hfag:ycp}
 }
 \end{figure}
The combined evidence for mixing is now compelling.  The probability 
for $x=y=0$ (no mixing) is excluded from a fit with no \CPV at 
the $9.8\sigma$ level.  The present data are also consistent with the 
absence of \CPV.  No evidence for \CPV\ in time-integrated decay 
rates for neutral $D$ mesons exists either, though the measurements 
are approaching the interesting level of sensitivity that could 
confront SM estimates.  Furthermore, systematic as well as statistical 
uncertainties are expected to shrink as more data at present and 
future flavour factories is accrued.

We should take this as a need to continue to press on with these
studies in case there is yet a hole in the SM at the energy scale that
can be probed with results from a flavour factory.

\begin{acknowledgments}
We are grateful for contributions from our colleagues at CLEO, Belle
and CDF for sharing their latest results.
The author wishes to thank \babar\ colleagues for providing excellent
results and much help in the preparation of this review.  This work
was supported by the US National Science Foundation under grant
number PHY 0757876.
\end{acknowledgments}


\begin{thebibliography}{10}

\bibitem{Pais:1975qs}
A.~Pais and S.~B. Treiman.
\newblock \CP\ Violation in Charmed Particle Decays.
\newblock {\em Phys. Rev.}, D12:2744--2750, 1975.

\bibitem{Aubert:2007wf}
B.~Aubert et~al.
\newblock Evidence for \DzDzb\ mixing.
\newblock {\em Phys. Rev. Lett.}, 98:211802, 2007.

\bibitem{Staric:2007dt}
M.~Staric et~al.
\newblock Evidence for \DzDzb\ Mixing.
\newblock {\em Phys. Rev. Lett.}, 98:211803, 2007.

\bibitem{Abe:2007rd}
K.~Abe et~al.
\newblock Measurement of \DzDzb Mixing in $\Dz\to\KS\pip\pim$ Decays.
\newblock {\em Phys. Rev. Lett.}, 99:131803, 2007.

\bibitem{Aaltonen:2007uc}
T.~Aaltonen.
\newblock Evidence for \DzDzb\ Mixing Using the CDF II Detector.
\newblock arXiv:0712.1567 [hep-ex], 2007.

\bibitem{Aubert:2007en}
B.~Aubert et~al.
\newblock {Measurement of $\DzDzb$ Mixing using the Ratio of Lifetimes for
  the Decays $\Dz\to\Km\pip$, $\Km\Kp$ and $\pim\pip$}.
\newblock {\em Phys. Rev.}, D78:011105, 2008.

\bibitem{Aubert:2008zh}
B.~Aubert et~al.
\newblock {Measurement of  \DzDzb\ Mixing from a Time-dependent Amplitude
  Analysis of $\Dz\to\Kp\pim\piz$ Decays}.
\newblock arXiv:0807.4544 [hep-ex], 2008.

\bibitem{Wolfenstein:1985ft}
Lincoln Wolfenstein.
\newblock {$\DzDzb$ Mixing}.
\newblock {\em Phys. Lett.}, B164:170, 1985.

\bibitem{Bigi:2000wn}
Ikaros I.~Y. Bigi and Nikolai~G. Uraltsev.
\newblock {$\DzDzb$ Oscillations as a Probe of Quark-hadron Duality}.
\newblock {\em Nucl. Phys.}, B592:92--106, 2001.

\bibitem{Golowich:2007ka}
Eugene Golowich, JoAnne Hewett, Sandip Pakvasa, and Alexey~A. Petrov.
\newblock Implications of $\DzDzb$ Mixing for New Physics.
\newblock {\em Phys. Rev.}, D76:095009, 2007.

\bibitem{Golowich:2006gq}
Eugene Golowich, Sandip Pakvasa, and Alexey~A. Petrov.
\newblock New physics contributions to the Lifetime Difference in $\DzDzb$ Mixing.
\newblock {\em Phys. Rev. Lett.}, 98:181801, 2007.

\bibitem{Grossman:2006jg}
Yuval Grossman, Alexander~L. Kagan, and Yosef Nir.
\newblock {New physics and CP violation in Singly Cabibbo Suppressed $D$ decays}.
\newblock {\em Phys. Rev.}, D75:036008, 2007.

\bibitem{Aitala:1996fg}
E.~M. Aitala et~al.
\newblock A Search for $\DzDzb$ Mixing and Doubly-Cabibbo-Suppressed Decays
  of the $\Dz$ in Hadronic Final States.
\newblock {\em Phys. Rev.}, D57:13--27, 1998.

\bibitem{Godang:1999yd}
R.~Godang et~al.
\newblock {Search for $\DzDzb$ Mixing}.
\newblock {\em Phys. Rev. Lett.}, 84:5038--5042, 2000.

\bibitem{Link:2004vk}
J.~M. Link et~al.
\newblock {Measurement of the Doubly Cabibbo Suppressed Decay $\Dz\to\Kp\pim$ and
  a Search for Charm Mixing}.
\newblock {\em Phys. Lett.}, B618:23--33, 2005.

\bibitem{Abe:2004sn}
K.~Abe et~al.
\newblock {Search for $\DzDzb$ Mixing in $\Dz\to\Kp\pim$ Decays and Measurement
  of the Doubly-Cabibbo-Suppressed Decay Rate}.
\newblock {\em Phys. Rev. Lett.}, 94:071801, 2005.

\bibitem{Aubert:2006rz}
B.~Aubert et~al.
\newblock Search for $\DzDzb$ Mixing in the Decays $\Dz\to\Kp\pim\pip\pim$.
\newblock arXiv:hep-ex/0607090, 2006.

\bibitem{Zhang:2006dp}
L.~M. Zhang et~al.
\newblock {Improved Constraints on $\DzDzb$ Mixing in $\Dz\to\Kp\pim$ Decays
  at Belle}.
\newblock {\em Phys. Rev. Lett.}, 96:151801, 2006.

\bibitem{hfag:ycp}
A~Schwartz et~al.
\newblock {Heavy flavour averaging group}.
\newblock {\em http://www.slac.stanford.edu/xorg/hfag/charm/ICHEP08/results_mixing.html}
\newblock 2008.

\bibitem{Liu:1994ea}
Tie-hui~(Ted) Liu.
\newblock {The $\DzDzb$ Mixing Search: Current Status and Future Prospects}.
\newblock arXiv:hep-ph/9408330, 1994.

\bibitem{Bergmann:2000id}
Sven Bergmann, Yuval Grossman, Zoltan Ligeti, Yosef Nir, and Alexey~A. Petrov.
\newblock {Lessons from CLEO and FOCUS Measurements of $\DzDzb$ Mixing
  Parameters}.
\newblock {\em Phys. Lett.}, B486:418--425, 2000.

\bibitem{Asner:2005sz}
D.~M. Asner et~al.
\newblock Search for $\DzDzb$ Mixing in the Dalitz Plot Analysis of 
  $\Dz\to\KS\pip\pim$.
\newblock {\em Phys. Rev.}, D72:012001, 2005.

\bibitem{phi:2008tp}
A~Zupanc et~al.
\newblock {Measurement of $\ycp$ in $D$ Meson Decays to \CP\ Eigenstates}.
\newblock 2008.

\bibitem{Aubert:2005sm}
B.~Aubert et~al.
\newblock {Dalitz Plot Analysis of $D^0 \to \bar{K}^0 K^+ K^-$}.
\newblock {\em Phys. Rev.}, D72:052008, 2005.

\bibitem{Aubert:2008bd}
B.~Aubert et~al.
\newblock {Improved Measurement of the CKM Angle Gamma in
  $B^{\pm}\to D^{(*)} K^{(*)\pm}
  Decays with a Dalitz Plot Analysis of $D$ Decays to $\KS\pip\pim$ and 
  $\KS\Kp\Km$}.
\newblock {\em Phys. Rev.}, D78:034023, 2008.

\bibitem{Buccella:1994nf}
F.~Buccella, Maurizio Lusignoli, G.~Miele, A.~Pugliese, and Pietro Santorelli.
\newblock {Nonleptonic Weak Decays of Charmed Mesons}.
\newblock {\em Phys. Rev.}, D51:3478--3486, 1995.

\bibitem{Bianco:2003vb}
S.~Bianco, F.~L. Fabbri, D.~Benson, and I.~Bigi.
\newblock {A Cicerone for the Physics of Charm}.
\newblock {\em Riv. Nuovo Cim.}, 26N7:1--200, 2003.

\bibitem{Petrov:2004gs}
Alexey~A. Petrov.
\newblock {Hunting for \CP\ Violation with Untagged Charm Decays}.
\newblock {\em Phys. Rev.}, D69:111901, 2004.

\bibitem{Aubert:2007if}
B.~Aubert et~al.
\newblock {Search for \CP\ Violation in the Decays $\Dz\to\Km\Kp$ and
  $\Dz\to\pim\pip$}.
\newblock {\em Phys. Rev. Lett.}, 100:061803, 2008.

\bibitem{Arinstein:2008zh}
K.~Arinstein.
\newblock {Measurement of the Ratio $B(\Dz\to\pip\pim\piz)/B(\Dz\to\Km\pip\piz)$
  and the Time-integrated CP Asymmetry in $\Dz\to\pip\pim\piz$}.
\newblock {\em Phys. Lett.}, B662:102--110, 2008.

\bibitem{cpv:2008rx}
M.~Staric et~al.
\newblock {Measurement of CP Asymmetry in Cabibbo Suppressed $\Dz$ Decays}.
\newblock arXiv:0807.0148, 2008.

\bibitem{Aubert:2008yd}
B.~Aubert et~al.
\newblock {Search for $\CP$ Violation in Neutral $D$ Meson Cabibbo-Suppressed
  Three-body Decays}.
\newblock {\em Phys. Rev.}, D78:051102, 2008.

\bibitem{cpv:2008zi}
P.~Rubin et~al.
\newblock {Search for \CP\ Violation in the Dalitz-Plot Analysis of
  $D^\pm\to K^+K^-\pi^\pm$}.
\newblock {\em Phys. Rev.}, D78:072003, 2008.

\bibitem{Schwartz:2008wa}
A.~J. Schwartz.
\newblock {Measurements of $\Dz\Dzb$ Mixing and Searches for \CP\ Violation: HFAG
  Combination of All Data}.
\newblock arXiv:0803.0082, 2008.

\end{thebibliography}


\end{document}